\documentclass[modern,dvipsnames]{aastex}

\shorttitle{MHD evolution of AR 11515}
\shortauthors{Kumar et al.}

\begin{document}

\title{Magnetohydrodynamics evolution of three-dimensional magnetic null in NOAA active region 11515 initiated using non-force-free field extrapolation}

\author{Sanjay Kumar} 
\affil{Department of Physics, 
Patna University, Patna 800005, India}

\author{Avijeet Prasad} 
\affil{Rosseland Centre for Solar Physics, University of Oslo, Postboks 1029 Blindern, 0315 Oslo, Norway} 
\affil{Institute of Theoretical Astrophysics, University of Oslo, Postboks 1029 Blindern, 0315 Oslo, Norway}

\author{Ranadeep Sarkar} 
\affil{Department of Physics, University of Helsinki, Helsinki 00560, Finland}

\author{R. Bhattacharyya}
\affil{Udaipur Solar Observatory, Physical Research Laboratory,
Dewali, Bari Road, Udaipur 313001, India}


\begin{abstract}
Magnetohydrodynamics simulation of active region NOAA 11515 is performed to examine the initiation of the M5.6 flaring event that starts around 10:43 UT on 2012 July 2. The simulation is conducted using an extrapolated non-force-free magnetic field generated from the photospheric vector magnetogram of the active region as the initial magnetic field. The magnetic field shows the presence of a three-dimensional (3D) magnetic null with the corresponding dome overlying a filament and a low-lying magnetic flux rope, observed in 304~\AA~ and 131~\AA~ respectively. 
 The simulated dynamics, triggered by the initial Lorentz force, lead to the bifurcations of the flux rope, which is similar to the observed bifurcation in the 131 \AA~ brightenings. Additionally, the rope exhibits a rise and reconnects at the 3D null. These reconnections convert field lines of the rope into the anchored outer spine of the 3D null --- explaining the occurrence of a nearby confined C-class flare. Further, the results show that the field lines of the flux rope reach the vicinity of the filament and become non-parallel to the field lines of the filament. This initiates the reconnections between the rope and the field lines of the filament --- activating the filament for the eruption. This interesting interaction of the flux rope and filament seems to contribute to the onset of the M-class flare.  
\end{abstract}


\section{Introduction}
Solar flares are the manifestation of the abrupt and explosive release of energy in the solar atmosphere. The flaring events generally relax the stressed coronal magnetic fields by releasing the stored magnetic energy {\citep{aschwanden}}. The coronal magnetic field lines (MFLs) being rooted in the solar surface with high plasma-$\beta$ are stressed by the flows at the surface and, consequently store the energy \citep{2011LRSP....8....6S, priest2014book}. However, the physics of the underlying processes which trigger the onset of the flaring events and the associated energy release is not fully understood \citep{2011LRSP....8....6S, priest2014book}. In this direction, the magnetic reconnection (MR) --- a diffusive process in which magnetic energy stored in the plasma gets converted into kinetic energy and heat accompanied by a change in magnetic topology ---  is widely accepted to be the physical process central to the sudden energy release  \citep{2011LRSP....8....6S, priest2014book}. 

The three-dimensional (3D) magnetic nulls (the locations where the magnetic field ${\bf{B}}=0$) are considered to be the potential sites for hosting the reconnections  {\citep{2005PhPl...12e2307P, hachami-2010, kumar&bhattacharyya2016phpl}}. In the solar atmosphere, the 3D null is characterized by its spine and dome-shaped fan structures resulting from the avoidance of the ${\bf{B}}=0$ location by the MFLs. As a consequence of the spine-fan structure, reconnections naturally commence at the 3D nulls  \citep{pontin-2013, pontin-2016, prasad+2018apj, 2020ApJ...892...44N}. Notably, recent magnetohydrodynamic (MHD) simulations document the occurrence of circular flare ribbons and blowout coronal jets \citep{masson+2009apj, 2012ApJ...760..101W, nayak+2019apj, liu-2020, prasad_2020} along with confined flares due to the reconnections at the null-point \citep{2007ApJ...662.1293U, prasad+2018apj}. In addition to the 3D nulls, also important are the magnetic flux ropes (MFRs), the magnetic structures  made of MFLs winding around a common axis {\citep{priest2014book}}.  Under favorable conditions, the flaring events are believed to be initiated by the evolution of the MFRs, which is governed by the reconnections    {\citep{inoue+2015apj, kumar+2016apj, prasad_2020}}.

To construct the suitable coronal magnetic field, the nonlinear-force-free-fields (NLFFFs) extrapolations are widely used 
which correspond to zero Lorentz force equilibrium state \citep{wiegelmann2008jgra,wiegelmann&sakurai2012lrsp,duan+2017apj}. Recently, magnetohydrodynamic (MHD) simulations initiated with the NLFFF extrapolations successfully explained 
the various eruptive events in solar corona \citep{jiang+2013apjl,kliem+2013apj,amari+2014nat, inoue+2014apj,inoue+2015apj,inoue2016peps}.
However, the NLFFF extrapolations are less accurate near the photosphere where plasma is non-force-free. In the photosphere, where the vector magnetograms are taken, the plasma-$\beta$ is non-negligible and approaches to unity \citep{gary2001soph}, so that the Lorentz force can not be neglected.  To tackle the issue, a technique called ``preprocessing" is used on the photospheric vector magnetograms that minimize the force and provide a suitable boundary condition for the NLFFF extrapolations \citep{wiegelmann2008jgra,wiegelmann&sakurai2012lrsp}. Recently, non-force-free-fields (NFFFs) extrapolation {\citep{hu&dasgupta2008soph, hu+2010jastp}} has emerged as a viable alternative to the NLFFF extrapolations. 
In the NFFFs extrapolations, the extrapolated field supports non-zero Lorentz force, which is crucial for the spontaneous onset of evolution. 
 Utilizing the NFFF extrapolations, recent MHD simulations successfully simulated coronal dynamics leading to solar flares, coronal jets, and coronal dimmings {\citep{prasad+2018apj, nayak+2019apj, prasad_2020, Bora_2022}}.

In this paper, we aim to explore the magnetohydrodynamics evolution of NOAA AR 11515 to understand the onset of an  eruptive M5.6 flare on 2012 July 2 at 10:43 UT, which is suggested to be triggered by a nearby confined C2.9 flare that occurred at 10:37 UT \citep{Louis_2014}. 
Importantly, the presented MHD simulation initiated by the NFFF extrapolation  provides a plausible explanation of the occurrence of the C-class flare and the  subsequent initiation of the M-class flare through the interactions of a pre-existing 3D null and the flux ropes.

The rest of the paper is arranged as follows: Section 2 briefly mentions the important aspects of the flaring event and the details of the initial non-force-free extrapolated field. Section 3 presents the results of the simulation along with their relation to the multi-wavelength observations. Important findings of the paper are summarized in  Section 4.

\section{Event observations and extrapolation}
\subsection{Flares in the NOAA AR 11515 on 2012 July 2}\label{sec:observations}
\begin{figure}[ht!]
\begin{centering}
\includegraphics[width=17cm]{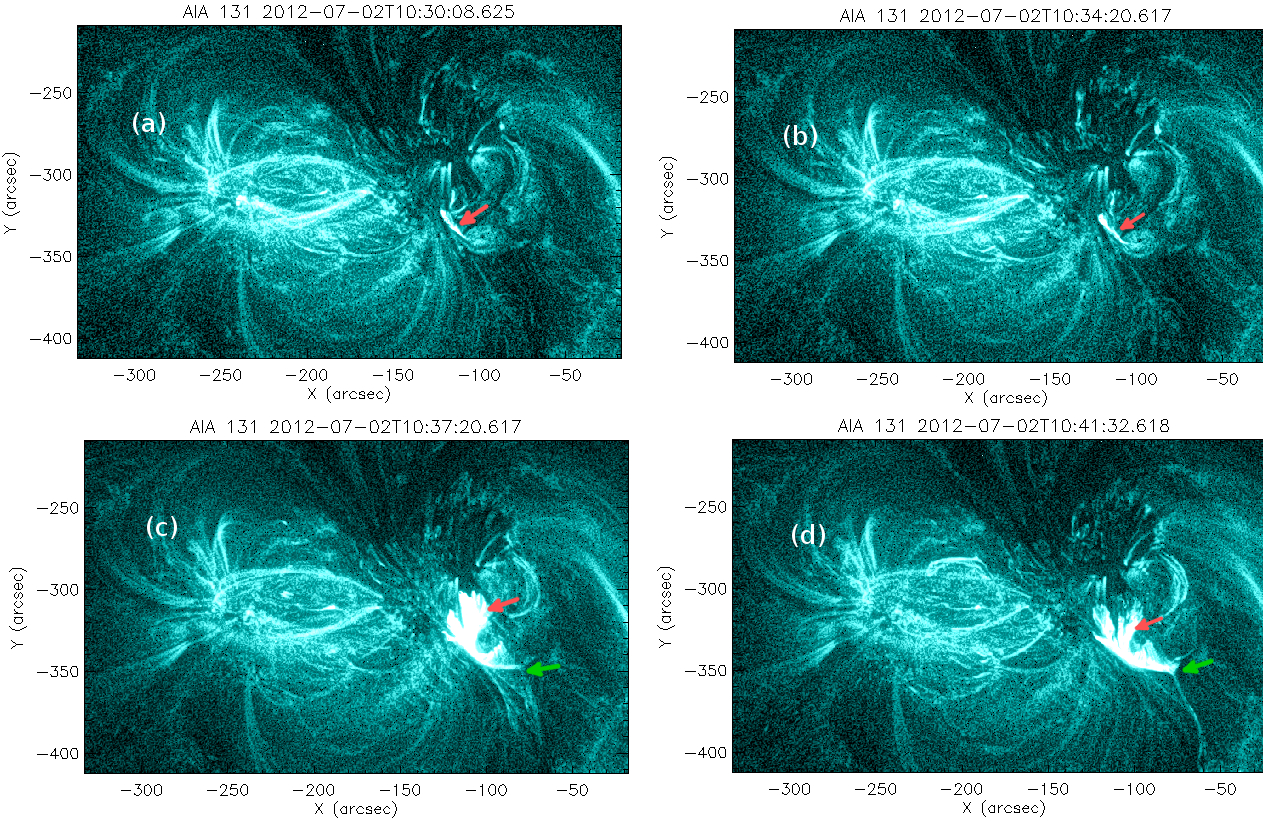}
\end{centering}
\caption{The appearance of the brightenings and its subsequent bifurcations in 131~\AA~are marked by red arrows in panels (a)-(d). Also notable is the faint southward spray of plasma (marked by green arrow in panels (c)-(d)).   }
\label{f1a-observations}
\end{figure}

\begin{figure}[ht!]
\begin{centering}
\includegraphics[width=17cm]{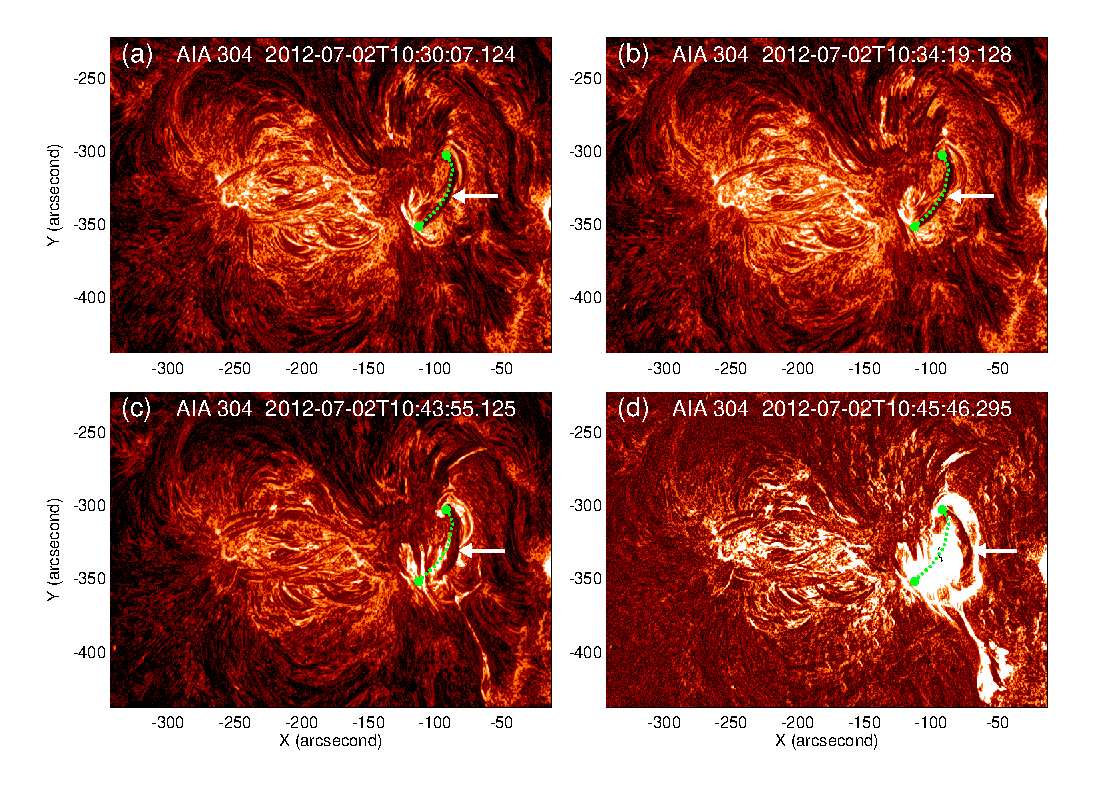}
\end{centering}
\caption{Panels (a)-(d) show the pre-existing filament and its subsequent evolution (marked by white arrows) in 304~\AA. The green-dashed line marks the location of filament at $10:30$ UT. Important is the activation of the filament around 10:43 UT (panel (c)) and subsequent  onset of filament eruption around 10:45 UT (panel (d)).   
}
\label{f1b-observations}
\end{figure}

The eruptive M5.6 flare occurs in NOAA AR 11515 on 2012 July 2. The various observational aspects of the flare are already studied by \citet{Louis_2014}. The temporal profile of GOES soft X-ray flux indicates that the flare starts at around 10:43 UT and peaks at $\approx$ 10:52 UT (not shown). Also notable is a preceding confined C2.9 flare in the vicinity of the M5.6 flare that initiated at 10:33 UT and peaked at 10:37 UT \citep{Louis_2014}. Here, in Figures \ref{f1a-observations} and \ref{f1b-observations}, we highlight a few important multi-wavelength observations of the flares obtained from the Atmospheric Imaging Assembly (AIA) \citep{lemen-2012} on board the Solar Dynamic Observatory (SDO). 
Figure \ref{f1a-observations} illustrates the dynamics of the AR during the precursor and impulsive phase of the M5.6 flare as observed in SDO/AIA 131~\AA~images. Panel (a) marks the presence of brightenings that extends in the north-east to south-west direction (marked by the red arrow) at around 10:30 UT. Typically, such brightenings are associated with the existence of a magnetic flux rope {\citep{joshi_2017, prasad_2020}}. Further, the rope maintains its structure until around 10:34 UT (panel (b)) and shows bifurcations in its eastern part at around 10:37 UT (marked by the red arrow in panels (c) and (d)), which is possibly related to the occurrence of the C2.9 flare.  The panels (c) and (d) also show the flare to be associated with a faint, southwards propagating spray (marked by the green arrows), without any eruptive behaviour \citep{Louis_2014}.

Figure \ref{f1b-observations} depicts the presence of a pre-existing filament (marked by a white arrow) at around 10:30 UT, which almost remains in a static situation until around the onset time of the M5.6 flare. The filament starts to rise and expand around 10:43 UT onwards (panel (c)) --- pointing towards the activation of the filament. The activation is then followed by the onset of the filament eruption (panel (d)). The erupting filament possesses close spatial and temporal associations with the M-class flare, indicating that the filament eruption possibly triggers the impulsive phase of the flare {\citep{Louis_2014}}. Moreover, \citet{Louis_2014} suggested the onset of the M-class flare to be triggered by the interactions of the filament with the nearby brightening (flux rope) associated with the preceding C-class flare, which points towards a causal connection between the flux rope dynamics and the filament eruption. 

\subsection{Extrapolated coronal magnetic field of AR 11515}\label{sec:nfff}
\begin{figure}[ht!]
\begin{centering}
\includegraphics[width=17cm]{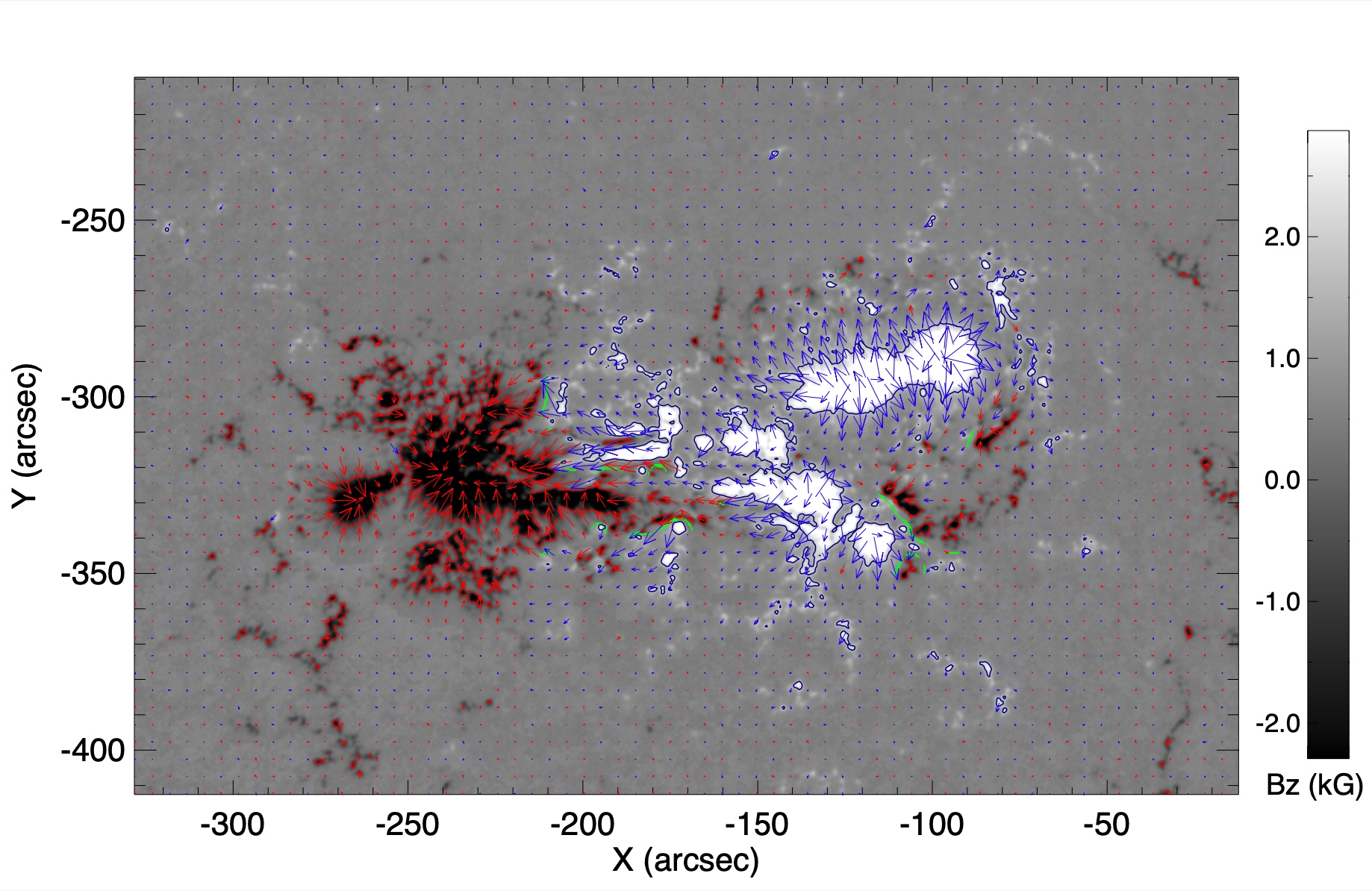}
\end{centering}
\caption{Figure illustrates HMI vector magnetogram of AR 11515 with start time (DATE\_OBS) of 10:34 UT on 2012 July 2. The red and blue arrows show the strength and direction of the transverse magnetic field, while green line mark PIL. }
\label{f2:magnetogram}
\end{figure}

We utilize the non-force free extrapolation model developed by \citet{hu&dasgupta2008soph,hu+2008apj,hu+2010jastp} to extrapolate the coronal magnetic field of AR 11515 at 10:34 UT on 2012 July 2 corresponding 
 to the photospheric vector magnetogram shown in Figure \ref{f2:magnetogram}. 
 Relevantly, the magnetogram is obtained from the Helioseismic Magnetic Imager \citep[HMI;][]{schou+2012soph} on board the Solar Dynamic Observatory (SDO). The magnetogram is taken from the `hmi.sharp\_cea\_720s' data series.
This data series provides the magnetic field on a  Cartesian grid which  is initially remapped onto a Lambert cylindrical equal-area (CEA) projection and then transformed into heliographic coordinates  \citep{bobra_2014}. We further crop the field of view to the region of interest to obtain a cutout of 640$\times$416 pixels in the $x$ and $y$ directions respectively. Noticeably, the cropping maintains the flux balance. In Figure \ref{f2:magnetogram}, the dark green line shows the polarity inversion line (PIL). It is noteworthy that, in the absence of significant dynamical changes from 10:30 UT to 10:34 UT in the multiwavelength observations of the events (Figures~\ref{f1a-observations} and \ref{f1b-observations}), we select the magnetogram at 10:34 UT for the extrapolation to reduce the computational cost of the MHD simulation.

\begin{figure}[ht!]
\begin{centering}
\includegraphics[width=17cm]{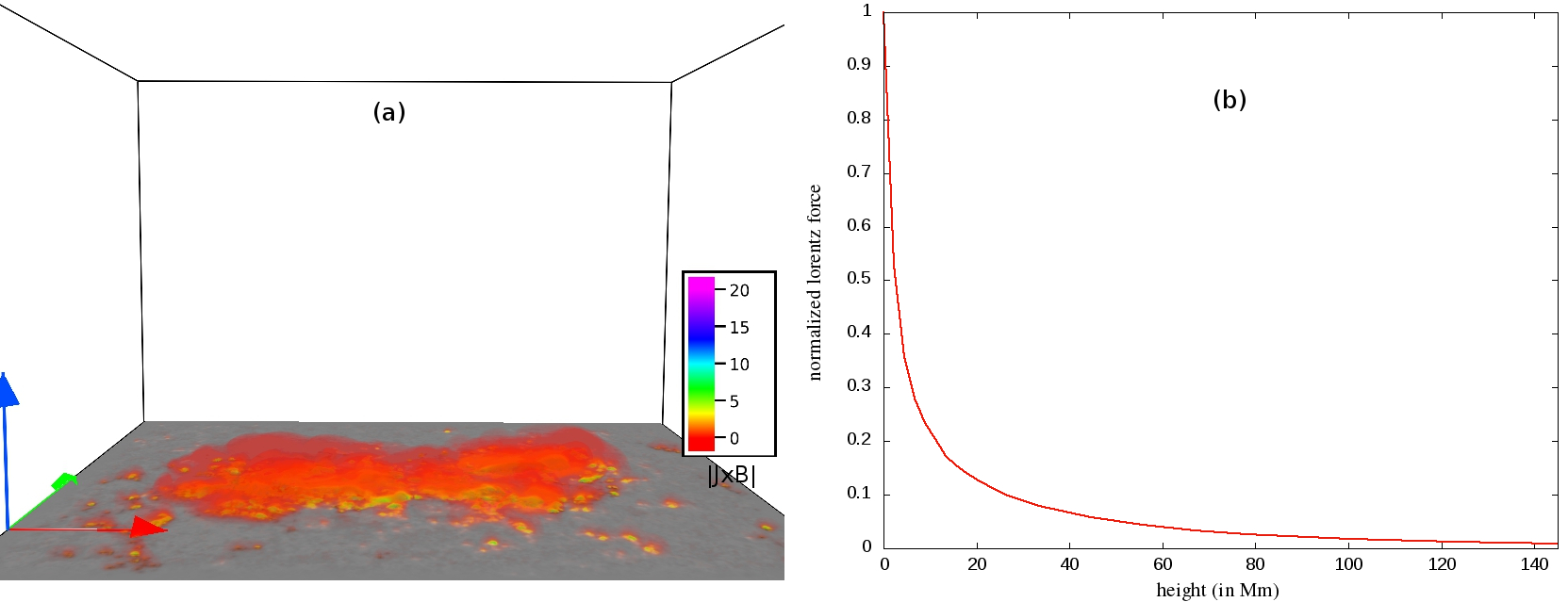}
\end{centering}
\caption{Panel (a) shows the distribution of the magnitude of the initial Lorentz force density in the computational domain. 
Panels (b) depicts the variation of magnitude 
of the horizontally averaged Lorentz force density with height (Mm). The force density is normalized with respect to its maximum value, as our focus is on its decay rate with height. Both the panels clearly illustrate the exponential decrease of the force density with height.}
\label{lorentz-force}
\end{figure}

\begin{figure}[ht!]
\begin{centering}
\includegraphics[width=17cm]{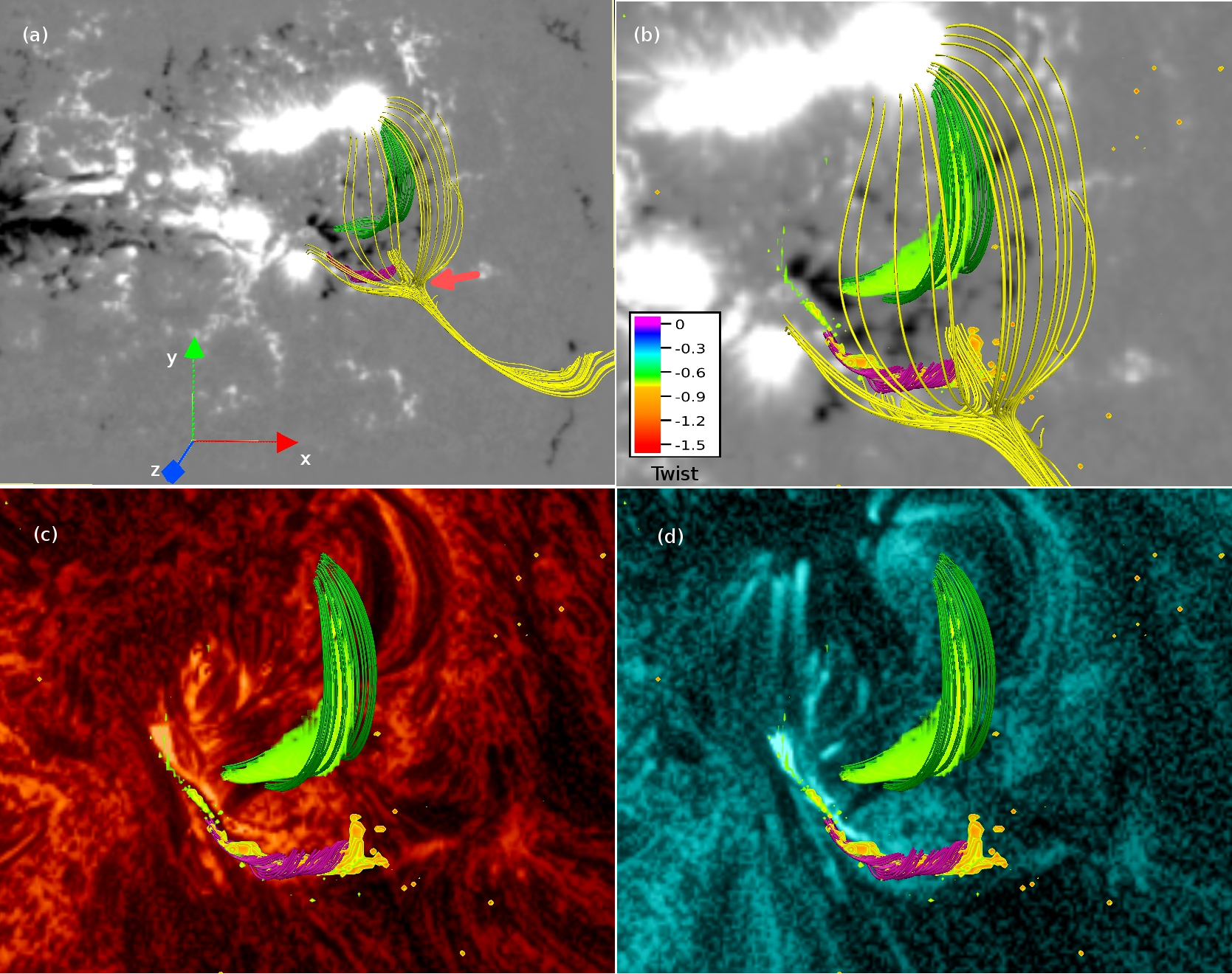}
\end{centering}
\caption{Panels (a) and (b) depict the zoomed-in view of the  extrapolated MFLs  with the magnetogram as the bottom boundary.
The location of the 3D null is marked by a red arrow in panel (a). 
The yellow MFLs represent the dome-shaped fan surface and spine axis of the 3D null. Panel (b) is further overlaid with the twist parameter --- showing the existence of two twisted magnetic structures below the dome. Panel (c) plots the structures with 304~~\AA~~ image of around 10:34 UT at the bottom boundary. The filament is almost co-located with one of the twisted structures plotted in green.  Panel (d) is overplotted with 131~~\AA~~ image of around 10:34 UT at the bottom boundary --- indicating the similar location of the magenta-colored twisted structure and the observed brightenings in  131~~\AA. }
\label{f3:extrapolation}
\end{figure}

The non-force-free extrapolation model is based on the principle of minimum energy dissipation rate, according to which, a plasma system prefers to relax toward a state which has a minimum dissipation rate {\citep{bhattacharyya+2007soph}}. The magnetic field ${\bf{B}}$ is then shown to satisfy the double-curl Beltrami equation {\citep{bhattacharyya+2007soph}}. The field ${\bf{B}}$ then can be written as the superposition of two linear-force-fields and a potential field \citep{hu&dasgupta2008soph}. Consequently, the Lorentz force associated with the field is non-zero. An iterative approach based on the minimization of the average deviation between the observed and the calculated transverse field on the photospheric boundary is then employed to get the extrapolated NFFF \citep{hu+2010jastp}.

Notably, in order to minimize the computational cost, the original domain with the spatial extension of $640\times 416$ pixels is re-scaled to $320\times 208$ pixels grid in $x$ and $y$ directions, respectively. The vertical extension of the domain is taken as $208$ pixels. We have checked that the magnetic structures remain preserved after the re-scaling by comparing the extrapolated field of the original magnetogram. 
Panel (a) of Figure $\ref{lorentz-force}$  illustrates the direct volume renderings of the Lorentz force density, which shows the presence of the Lorentz force at lower heights. To further confirm, in panel (b) of the figure, we plot the variation of horizontally averaged Lorentz force density with height. The averaged Lorentz force density falls
off sharply with height. Hence, Figure $\ref{lorentz-force}$  suggests that the Lorentz force mostly resides near the photosphere and almost vanishes at coronal heights, similar to the typical description of the solar corona. 
The averaged value of the Lorentz force associated with the extrapolated field at the bottom boundary is of the order of $10^{22}$ dyne. Interestingly, a recent observational study by \citet{Sarkar_2019} has estimated the change in the Lorentz force during flaring events, which is approximately $(2-5) \times 10^{22}$ dyne.
In our simulation, the non-zero Lorentz force at lower heights plays a crucial role in initiating the simulated dynamics of the active region that can lead to the flaring events.

The magnetic field lines of the extrapolated field are plotted in Figure \ref{f3:extrapolation}. In this and subsequent figures, the $x$, $y$, and $z$ axes of the computational domain are represented by red, green, and blue axes, respectively. The figure documents the existence of a pre-existing 3D null (marked by a red arrow in Fig. \ref{f3:extrapolation}(a)) and, the corresponding spine axis and the dome-shaped fan surface are represented by the yellow field lines. Notable are the field lines of the outer spine (extends in the southward direction) to be anchored at the bottom boundary (panel (a)). Further important is the presence of two twisted magnetic structures (having co-located high values of twist parameter), residing under the dome of the 3D null (see panel (b) of Figure \ref{f3:extrapolation}). Relevantly, the twist parameter measures the twist number of a field line and is calculated by integrating field-aligned current $\bf{J} \cdot \bf{B}/|B|^2$ along a field line \citep{Berger_2006, 2016ApJ...818..148L}. 
The maximum magnitude of the twist parameter for the magenta-colored magnetic structure is close to 1.5, while the maximum magnitude of the parameter is around 1.1 for the green-colored structure. This suggests that these structures represent magnetic flux ropes {\citep{prasad_2020}}.
Notably, the flux rope presented by the green field lines is almost co-spatial with the filament observed in 304~\AA~ (Fig. \ref{f3:extrapolation}(c)) --- indicating the rope to be the magnetic structure of the filament. Furthermore, the other rope plotted by magenta field lines has a higher magnitude of the twist parameter in comparison to the green-colored rope and co-locates with the brightenings observed in 131~\AA~ around 10:34 UT (Fig. \ref{f3:extrapolation}(d)). Due to the higher twist, the magenta-colored rope supports a larger electrical current and, therefore, seems to be responsible for the brightenings in 131~\AA.

\section{Magnetohydrodynamics evolution of AR 11515}
\subsection{Governing MHD equations}
The evolution of AR 11515 is simulated by numerically solving equations of magnetohydrodynamics in implicit large eddy simulation mode \citep{grinstein2007book}.  With a focus on exploring the changes in field line topology, here we consider the incompressible, thermally inactive plasma with zero physical resistivity  \citep{bhattacharyya+2010phpl,kumar+2014phpl,kumar+2015phpl}. As the involved magnetic field is sufficiently strong, the magnetic pressure dominates over other pressures such as thermal, gravitational, and therefore, their negligence is justifiable. The set of dimensional-less MHD equations is then given as: 

\begin{eqnarray}
\label{stokes}
&  \frac{\partial{\bf{v}}}{\partial t} 
+ \left({\bf{v}}\cdot\nabla \right) {\bf{ v}} =-\nabla p
+\left(\nabla\times{\bf{B}}\right) \times{\bf{B}}+\frac{\tau_a}{\tau_\nu}\nabla^2{\bf{v}},\\
\label{incompress1}
&  \nabla\cdot{\bf{v}}=0, \\
\label{induction}
&  \frac{\partial{\bf{B}}}{\partial t}=\nabla\times({\bf{v}}\times{\bf{B}}), \\
\label{solenoid}
 &\nabla\cdot{\bf{B}}=0.
\end{eqnarray}

\noindent The magnetic field strength ${\bf{B}}$ and the plasma velocity ${\bf{v}}$ are normalized by the average magnetic field strength ($B_0$) and  the Alfv\'{e}n speed
($v_a \equiv B_0/\sqrt{4\pi\rho_0}$ with $\rho_0$ representing the constant mass density), respectively. The plasma pressure $p$, the spatial-scale $L$, and the temporal scale $t$ are normalized by ${\rho {v_a}^2}$, length-scale of the vector magnetogram ($L_0$), and the Alfv\'{e}nic transit time ($\tau_a=L_0/v_a$), respectively. 
Here $\tau_\nu$ represents viscous diffusion time scale 
($\tau_\nu= L_0^2/\nu$), with $\nu$ being the kinematic viscosity. The pressure $p$
satisfies the elliptic boundary value problem, 
generated by imposing the incompressibility constraint (\ref{incompress1}) on the momentum transport equation (\ref{stokes}) (see \citep{bhattacharyya+2010phpl}).

The MHD equations are solved using the well-established numerical model EULAG-MHD \citep{smolarkiewicz&charbonneau2013jcoph}. The details of the model are described in \citet{smolarkiewicz&charbonneau2013jcoph} and references therein. Importantly, in the absence of the physical magnetic diffusivity (\ref{induction}), the dissipative property of the model, intermittently and adaptively, regularizes the under-resolved scales by simulating magnetic reconnections. Notably, in our previous works \citep{prasad+2018apj, nayak+2019apj, prasad_2020, Bora_2022}, we successfully simulated the dynamics of various active regions with the model and explained the solar transients such as flares, coronal jets, circular brightenings in the active regions.

It is noteworthy that, in the simulation, the initial velocity is taken to be $\bf{v}=0$. The dynamics is then developed by the Lorentz force associated with the initial extrapolated field (see Figure \ref{lorentz-force}). The direct volume renderings of the Lorentz force density (Figure \ref{lorentz-force}(a)) shows the presence of Lorentz force at lower heights and the 
higher values of the magnitude of the force are cospatial  with the strong-field regions (having high values of $B_z$). This Lorentz force existing at lower height initiates the simulated evolution by pushing the plasma from its initial stationary state and generates plasma flow. The field lines being frozen into the plasma get deformed by the plasma flow which, modify the initial distribution of the Lorentz force. Moreover, once the flow is generated the other forces such the pressure gradient and the viscous drag also come into play. Therefore, once initiated by the Lorentz force, the other forces also contribute to the simulated dynamics.  


\subsection{Simulated evolution}

\begin{figure}[ht!]
\begin{centering}
\includegraphics[width=17cm]{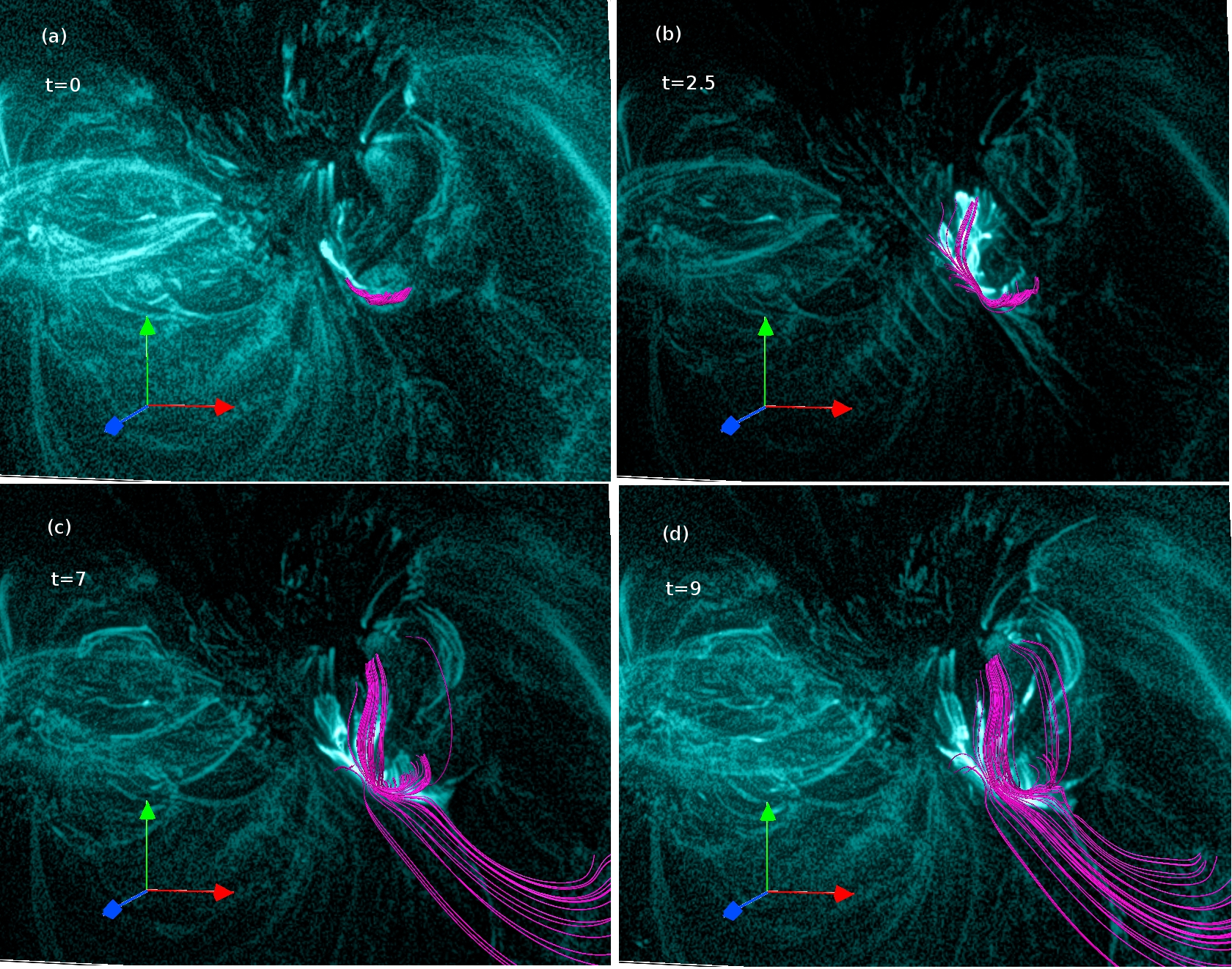}
\end{centering}
\caption{Evolution of the magenta-colored flux rope co-relating the brightenings in 131~\AA. The evolution is overplotted with the co-current 131~\AA~ images. Notable is the similar bifurcations of   the rope and brightenings in  131~\AA~ (panel (b)). Panels (c) and (d) indicates the rise of the rope and its reconnection at the 3D null.     }
\label{f4:flux-rope}
\end{figure}

The presented simulation is conducted in a computational domain having $320\times 208 \times 208$ grid points for a physical domain spanning $[0,1.54]\times [0,1]\times [0,1]$ units in $x$, $y$, and $z$, respectively, where an unit length approximately corresponds to $145$ Mm. The simulation is initiated with the NFFF extrapolated magnetic field as shown in Figure \ref{f3:extrapolation}. As mentioned above, with an initial motion-less state, the simulated dynamical evolution is generated by the initial non-zero Lorentz force associated with the NFFF.   The resulted flow is incompressible, an assumption also unitized by \citet{dahlburg+1991apj, aulanier+2005aa}. As our focus is to understand the topological changes responsible for the flare onset, the assumption seems to be justifiable in the tenuous coronal medium. $B_z$ is kept constant at the bottom boundary of the computational domain, because the change of magnetic flux at the boundary is minimal during the flares (not shown). Other boundaries are kept open  by continuing all the components of ${\bf{B}}$ to the boundaries in order to vanish the net magnetic flux passing through them {\citep{prasad+2018apj}}. 
The value of the dimensional-less constant $\tau_a / \tau_\nu$ is chosen to be $5 \times 10^{-4}$, which is around 20 times larger than its coronal value. The higher  $\tau_a / \tau_\nu$ for the simulation only speeds up the dynamical evolution and,  hence, reduces the computational cost, without any effect on the magnetic topology.  The spatial unit step $\Delta x = 0.0048$ and time step (normalized by the Alfv\'{e}n  transit time $\tau_a \sim 20s$) $\Delta t = 2\times10^{-3}$ are selected to satisfy the Courant-Friedrichs-Lewy (CFL) stability condition \citep{courant1967jrd}. The simulation is carried out for 1500 $\Delta t$. Because of the higher $\tau_a / \tau_\nu$, the simulated dynamics is expected to be around 20 times faster than the coronal dynamics and, therefore, we need to multiply 1500 $\Delta t$ by 20 to directly compare the simulation results with observations. 
As a result, the total simulation time roughly corresponds to an observation time of 20 Minutes. For a direct comparison of the simulated evolution with the observations, we present the time in units of $3.2 \tau_a \approx 1$ minute in describing the simulation results.

\begin{figure}[ht!]
\begin{centering}
\includegraphics[width=17cm]{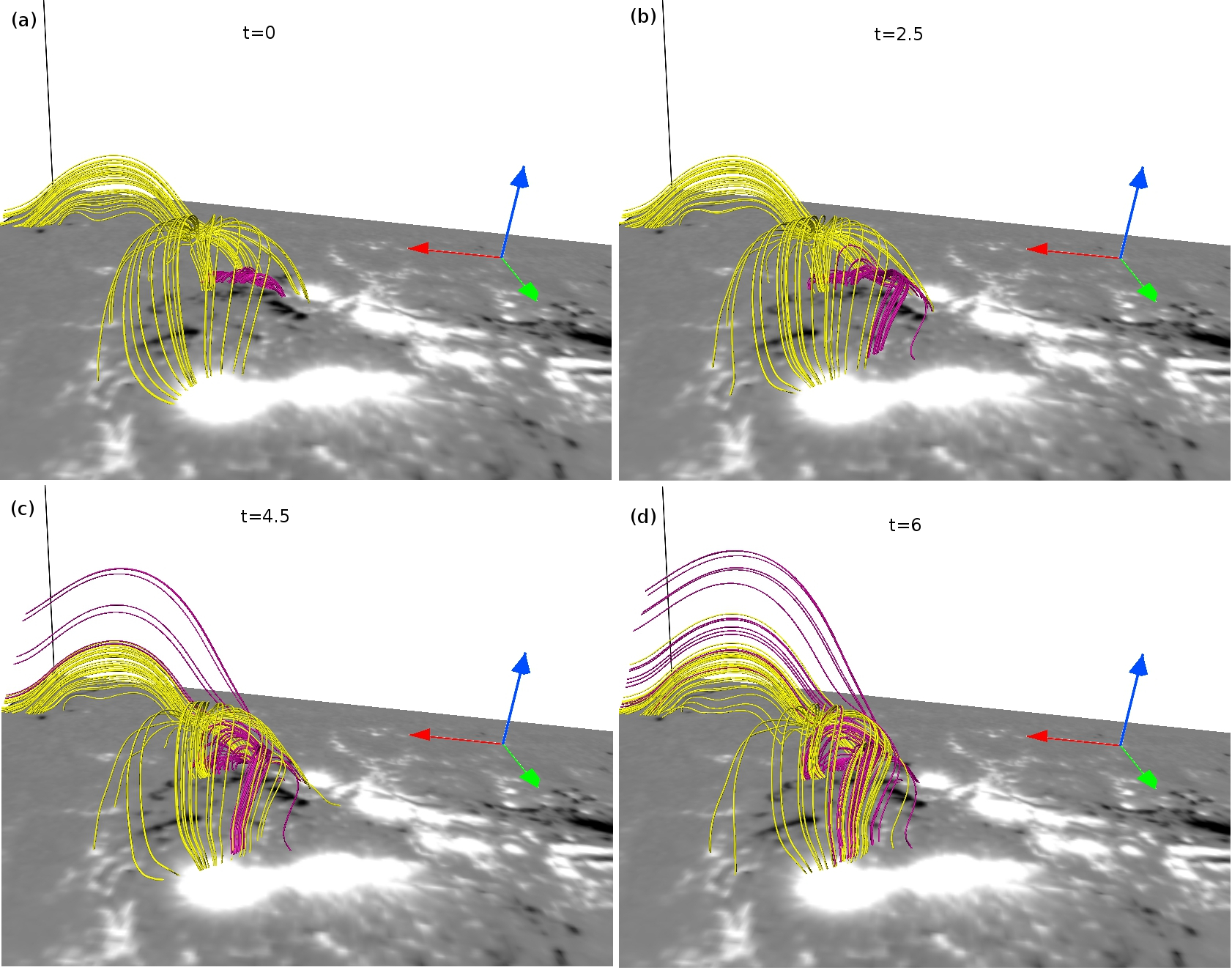}
\end{centering}
\caption{The figure depicts the evolution of the 3D null (in yellow) and the flux rope (in magenta).
The time is shown in units of $3.2 \tau_a \approx 1$ minute.
The rope located below the dome rises to the 3D null (panel (b)) and undergo magnetic reconnections (panels (c) and (d)). }
\label{f5:flux-rope-3D-null}
\end{figure}

To understand the dynamics of the pre-flare stage of the M-class flare, we first describe the evolution of the flux rope, correlating the brightenings in 131~\AA~(Fig. \ref{f3:extrapolation}(c)).  The evolution is shown in Figure \ref{f4:flux-rope}. In the figure, the bottom boundary is overlaid with concurrent 131~\AA~ images. Relevantly, for plotting the evolution of the magnetic field lines, in this and the subsequent figures, we utilize ``field line advection” technique in-built in the VAPOR visualization package \citep{clyne2005}, In the technique, one representative point for a selected field line is advected by the velocity field and, then the advected point is used as a seed to plot the field line at later time \citep{mininni2008}. For a detailed description of the technique and its successful illustration
in ideal as well as non-ideal magnetofluids, readers are referred to \citep{clyne2007, mininni2008}.
In Figure \ref{f4:flux-rope}, the flux rope exhibits the bifurcations, similar to the ones in the brightenings in 131~\AA~ (see Fig. \ref{f4:flux-rope}(b)) around at $t=2.5$, which approximately corresponds to 10:37 UT in the observations. In addition, the rope appears to rise and ultimately, reconnect at the 3D null (see Fig. \ref{f4:flux-rope}(c)-(d)).  To clearly illustrate the null-point reconnections of the flux rope, in Figure \ref{f5:flux-rope-3D-null}, we present the evolution of the rope (in magenta) overlaid with the field lines of the 3D null (in yellow). Important are the rise of the field lines of the rope toward the 3D null (Fig. \ref{f5:flux-rope-3D-null}(b)) and, their subsequent reconnections at the null (Fig. \ref{f5:flux-rope-3D-null}(c)). These reconnections repeat in time and allow the field lines of the rope to come out of the dome and transform into the field lines of the outer spine of the 3D null. The conversion of the rope-MFLs into the outer spine can transport the plasma ( confined in the rope) along the outer spine. However, the outer spine being anchored in the southward direction does not allow the plasma to escape from the corona --- explaining the observed faint southward brightenings during the C-class flare (see Fig. \ref{f1a-observations}(d) and Fig. \ref{f4:flux-rope}(d)). 
Hence, the bifurcations of the rope and its repeated reconnections at the 3D null seem to contribute to the occurrence of the C-class flare.

\begin{figure}[ht!]
\begin{centering}
\includegraphics[width=17cm]{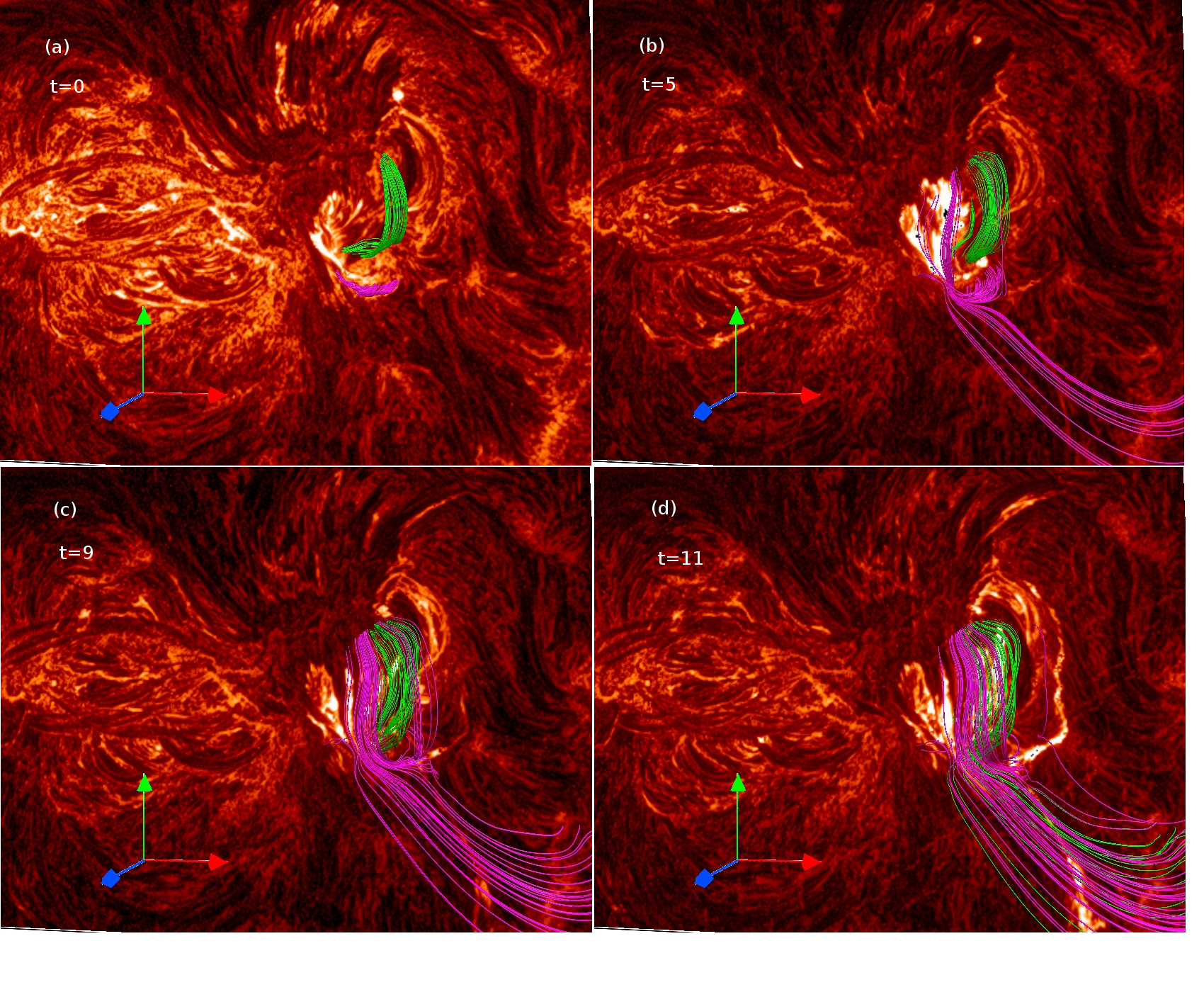}
\end{centering}
\caption{The figure shows the dynamics of the flux rope representing the filament (in green) along with the magenta-colored rope. The figure is overlaid with the 304~\AA~images. Notably, the green-colored rope shows a few dynamical changes until $t=9$ and starts reconnecting after that.   }
\label{f6:filament}
\end{figure}

To explore the dynamical evolution of the pre-existing filament, Figure \ref{f6:filament} illustrates the evolution of the flux rope (in green) representing the filament along with the magenta-colored flux rope. The figure is further overlaid with the concurrent 304~\AA~ images at the bottom boundary. Noticeably, the green-colored rope does not show much dynamical changes  until around $t=9$  (Fig. \ref{f6:filament}(a)-(c)) --- matching the observed behaviour of the filament until around 10:43 UT (Figure \ref{f1b-observations}). Subsequently the rope appears to undergo reconnections at around $t=11$, as indicated by the opening of the rope-MFLs (Fig. \ref{f6:filament}(d)).

\begin{figure}[ht!]
\begin{centering}
\includegraphics[width=17cm]{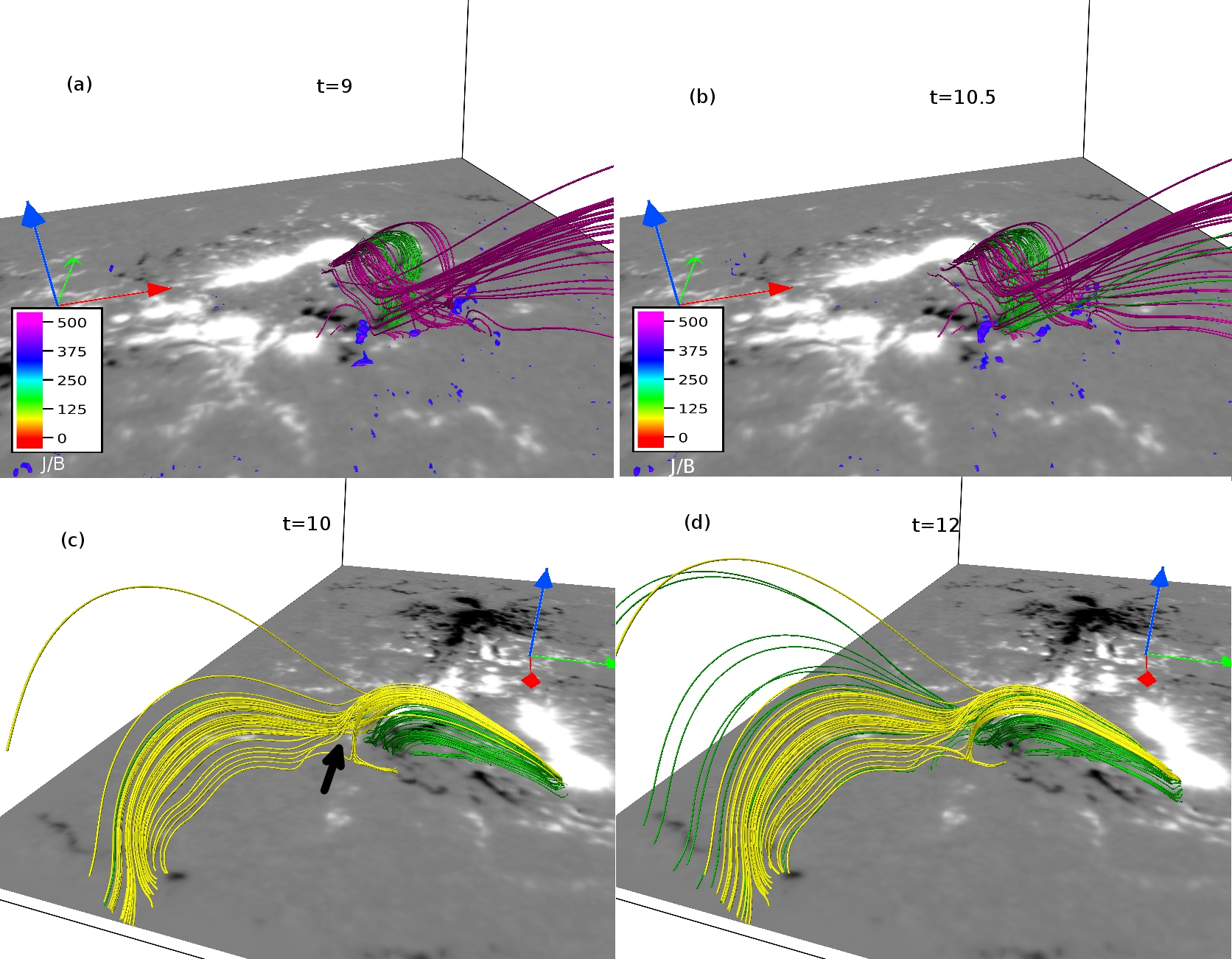}
\end{centering}
\caption{Panels (a) and (b) illustrate the evolution of the green-colored rope (filament) and the magenta-colored rope overplotted with $|\bf{J}|$/$|\bf{B}|$. Here, the time is in units of around $1$ minute.
Non-parallel field lines of these ropes reach in the close vicinity and develop high  $|\bf{J}|$/$|\bf{B}|$ regions, which explain the onset of reconnections in green-colored rope. 
Panels (c) and (d) shows the green-colored rope along yellow field lines of the 3D null (marked by black arrow in panel (c)) at $t=10$ and $t=12$. Green field lines are appear to move toward the 3D null and get reconnected.  }

\label{f8:filament}
\end{figure}

\begin{figure}[ht!]
\begin{centering}
\includegraphics[width=17cm]{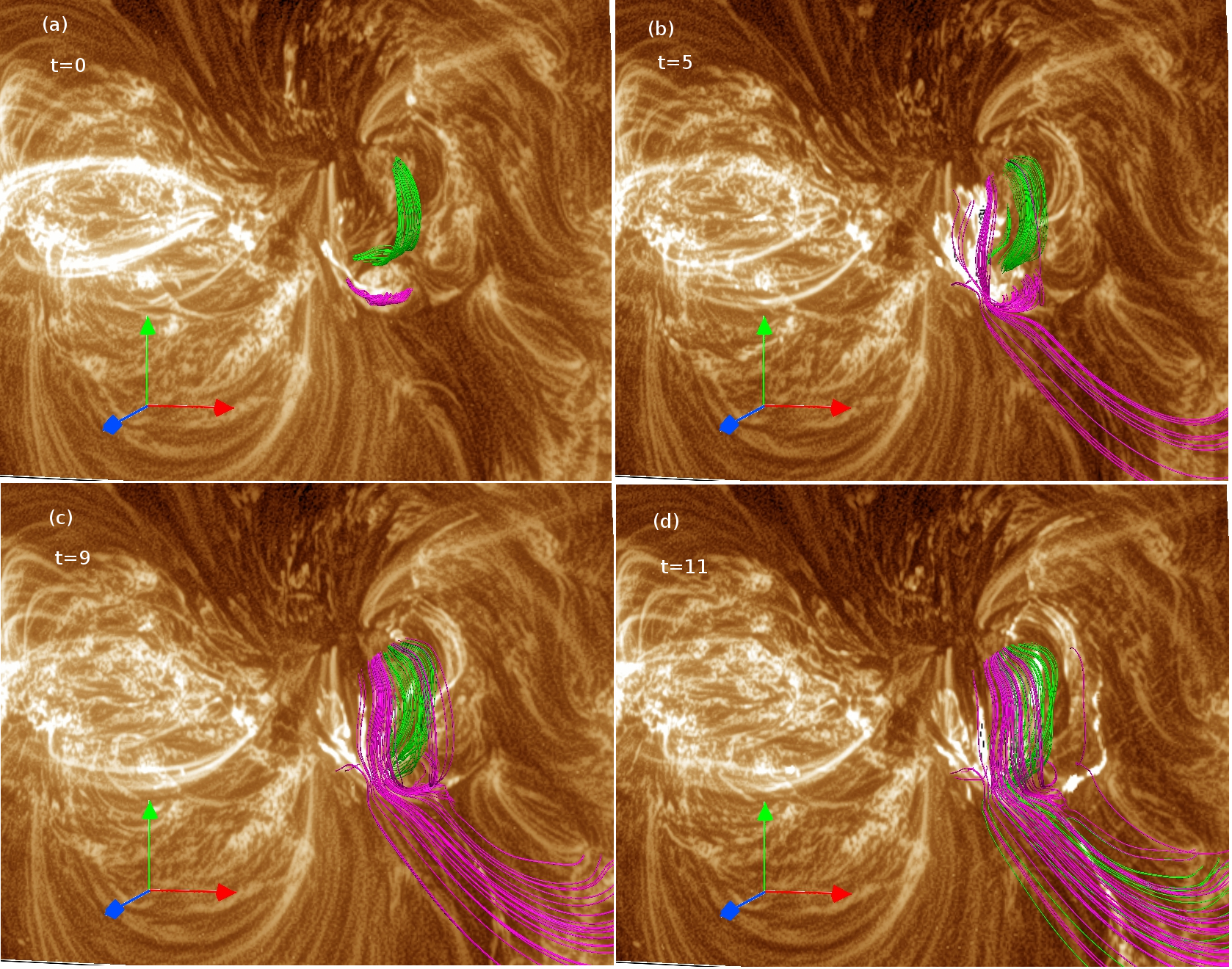}
\end{centering}
\caption{Here the dynamics of the filament associated rope (in green) and the magenta-colored rope (as in Figure \ref{f6:filament}) are overplotted with  the co-temporal 193~\AA~ observations. The post-reconnected field lines of the ropes match well with the bright structures extending along the right bottom in the observational images. }
\label{193-observations}
\end{figure}

To examine the reconnections of the rope associated with the filament, in panels (a) and (b) of Figure \ref{f8:filament}, we plot the evolution of the green-colored rope and the magenta-colored rope. The figure is overplotted with $|\bf{J}|$/$|\bf{B}|$. The evolution illustrates that  the green-colored rope and the magenta-colored rope (located below the dome of the 3D null) deform in such a way that the corresponding non-parallel MFLs move in close proximity. Consequently, the gradient of $\bf{B}$
steepens and  strong electric currents originate in the vicinity of
the non-parallel field lines at $t=9$, evidenced by the presence of high values of $|\bf{J}|$/$|\bf{B}|$. As a result, the scales become under-resolved, which initiates magnetic reconnections that repeatedly occur in
time, and are responsible for the opening-up of the green-colored rope (clearly identifiable in Fig. \ref{f8:filament}(b)). 
Similar to the magenta-colored rope evolution (Figure \ref{f5:flux-rope-3D-null}), the reconnected green field lines appear to become part of the longer outer spine field lines. 
Hence, the opening of the filament seems to be initiated by the reconnections between the parts of the ropes, which are located below the dome. Additionally, we also notice that the green-colored rope seems to move toward the 3D null (marked by black arrow in Fig. \ref{f8:filament}(c)) and starts to reconnect at the null (Fig. \ref{f8:filament}(d)).  Such reconnections can further contribute to the generation of the longer green outer spine field lines and speed up the filament eruption process.

To further support the occurrence of such repeated reconnections, in Figure \ref{193-observations}, we plot the evolution of filament-associated rope (in green) and the magenta-colored rope (same as Figure \ref{f6:filament}) overlaid with co-current SDO/AIA 193 \AA~ images. Noticeably, the bifurcation of the magenta-colored rope at $t=5$ matches the observed bifurcation in the brightenings (Fig. \ref{193-observations}(b)). Moreover, the panel (d) of the figure also suggests that there are bright features (situated at the right bottom in the observational 193 \AA~ images) corresponding to the post-reconnected semi-parallel field lines of the ropes (in green and magenta) --- suggesting a direct connection between the reconnections and the features.

Remarkably, the interactions of two ropes through reconnections
indicate that the occurrence of the C-class flare serves as the pre-cursor for the M-class flare. Because of the reconnections, the field lines of the green-colored rope associated with filament start to open up, which enable the otherwise confined cool plasma of the filament to escape along the post-reconnection opened field lines. Hence, the opening of the field lines through the reconnections can provide a potential explanation of the onset of the filament eruption that leads to the initiation of the M-class flare.

\section{Summary and Discussion}
The paper reports an MHD simulation of the initiation of an M5.6 flare in the NOAA AR 11515 on 2012 July 2. The focus of the paper is to understand the dynamics and evolution of the pre-existing 3D null and the magnetic flux ropes during the initiation phase of the flare. The initial magnetic field is obtained by extrapolating the photospheric vector magnetogram of the active region obtained from HMI/SDO at 10:34 UT on 2012 July 2 using the non-force-free extrapolation technique. The extrapolated field supports a Lorentz force at lower heights which becomes negligible in the corona --- matching the standard picture of the coronal magnetic field. The force, however, plays a crucial role in generating the self-triggered dynamical evolution from the initial static state. As the non-parallel MFLs approach in close proximity during the evolution, the under-resolved scales develop. The employed numerics regularize the scales with simulated magnetic reconnections by producing locally adaptive residual dissipation.

SDO/AIA multiwavelength observations (particularly in 131 \AA~ channel)  show the  signature of a flux rope and its bifurcations during the pre-flaring activities. Subsequent dynamics documents the rise of the rope and appearances of the faint brightenings extending along the southward direction --- indicating the reconnections of the rope at the 3D null, which appear to contribute to the occurrence  of a C-class flare in the vicinity of the M5.6 flare. A pre-existing filament is also observed in 304 ~\AA~ channel which initially remains almost stable and exhibits a sudden rising motion with the loss of its material --- pointing towards the initiation of the eruption during the onset of the M-class flare.

The non-force-free extrapolation is able to successfully capture the presence of a flux rope (represented by the highly twisted field lines), located under the magnetic topology of a 3D null point located close to the flaring region. Importantly, the flux rope  field lines explain the signature of the flux rope observed in the SDO/AIA 131 \AA~ channel. Another set of highly twisted field lines, implying the presence of a second flux rope, also exists which is almost co-spatial with the filament observed in the SDO/AIA 304 \AA~ channel.

The simulated MFL evolution documents the bifurcations of the flux rope co-related to the 131~\AA~ brightenings and its subsequent rise, which lead to the reconnections of the rope at the 3D null. The evolution explains the dynamical changes in the brightenings observed in the SDO/AIA 131 ~\AA~ channel during the C-class flare.  
The other flux rope associated with the filament shows a little changes in its early phase of evolution and then manifests reconnections. These reconnections are attributed to the movement of the non-parallel field lines of these two ropes in the close vicinity and the consequent development of high gradients in the magnetic field. These repeated reconnections allow to open-up the field lines of the filament and, ultimately trigger the eruption of the filament --- initiating the M-class flare.

Overall, the presented simulation successfully captures the dynamical evolution of the flux ropes and the 3D null, which leads to the precursor C-class flare and, ultimately initiates the M-class flare by activating the filament eruption. On the flip side, the simulated dynamics lack the significant eruption corresponding to the sudden and rapid rise of the flux rope, which is important to understand the consequent CME dynamics.  This can be attributed to the continuously operating viscous relaxation, which keeps on decaying the available free magnetic energy needed for generating the sudden rise. Additionally, the employed simplification  of the incompressibility and constant initial density may also play a role in the sudden rise of the flux rope.  Therefore, the simulation can be further advanced to simulate the eruption by relaxing the incompressibility and including an apt physical resistivity, which is kept for future assignments.

\section*{Conflict of Interest Statement}

The authors declare that the research was conducted in the absence of any commercial or financial relationships that could be construed as a potential conflict of interest.

\section*{Author Contributions}
SK, AP and RS contributed to the initial conception of the paper. The MHD simulation was conducted by AP and the analysis was carried out primarily by SK. The main draft was prepared by SK and, RB helped in the interpretation of the simulation results and the overall presentation of the paper. All authors participated in discussions and revisions on the draft.

\section*{Funding}
AP is supported from the Research Council of Norway through its Centres of Excellence scheme, project number 262622, as well as through the Synergy Grant number 810218 459 (ERC-2018-SyG) of the European Research Council. AP also obtains partial support from NSF award AGS-2020703. RS is supported from the project EFESIS (Exploring the Formation, Evolution and Space-weather Impact of Sheath-regions), a personal grant of RS (Academy of Finland Grant 350015).

\section*{Acknowledgments}
We acknowledge the use of the visualization software VAPOR (www.vapor.ucar.edu) for generating relevant graphics. Data and images are courtesy of NASA/SDO and the HMI and AIA science teams. SDO/HMI is a joint effort of many teams and individuals to whom we are greatly indebted for providing the data. The authors thank the reviewers for providing insightful comments and suggestions to enhance the scientific content as well as the presentation of the article.

\section*{Data Availability Statement}
The datasets generated for this study are available on request to the corresponding author.

\bibliographystyle{aasjournal}
\bibliography{ms}

\end{document}